\documentclass[12pt]{article}
\usepackage{epsfig,latexsym,amssymb,amsmath}
\usepackage{color}
\usepackage{graphicx}
\begin{document}

\begin{center}
{\large \bf Thermodynamic Product Relations for Generalized Regular Black Hole}
\end{center}

\vskip 5mm

\begin{center}
{\Large{Parthapratim Pradhan\footnote{E-mail: pppradhan77@gmail.com}}}
\end{center}

\vskip  0.5 cm

{\centerline{\it Department of Physics}}
{\centerline{\it Vivekananda Satavarshiki Mahavidyalaya}}
{\centerline{\it (Affiliated to Vidyasagar University)}}
{\centerline{\it Manikpara, West Midnapur}}
{\centerline{\it West Bengal~721513, India}}

\vskip 1cm

\begin{abstract}
We derive thermodynamic product relations for four-parametric regular black hole (BH) solutions 
of the Einstein equations coupled with a non-linear electrodynamics source. The four parameters 
can be described by the mass ($m$), charge ($q$), dipole moment ($\alpha$) and quadrupole moment 
($\beta$) respectively.  We study its complete thermodynamics. We compute different thermodynamic 
products i.e. area product, BH temperature product, specific heat product and Komar energy product
respectively. Furthermore, we show that some complicated function of horizon areas that is  
indeed \emph{mass-independent} and could turn out to be \emph{universal}.
%\keywords{Entropy product \and  Area product \and Ay\'{o}n-Beato and Garc\'{i}a Black Hole }
\end{abstract}

\section{Introduction}
It has been examined that the thermodynamic product for Reissner Nordstr{\o}m (RN) BH, Kerr BH and 
Kerr-Newman (KN) BH\cite{ah09}, a simple area product is sufficient to draw a conclusion that the product 
of area(or entropy) is an \emph{universal} quantity. The universal term is used here to 
describe when the product of any thermodynamic quantities is simply \emph{mass independent}. 
Alternatively mass-dependent thermodynamic quantities implies that they are \emph{not} an universal 
quantity. This is strictly follow throughout the work.

In case of  RN-AdS \cite{mv13}, Kerr-AdS, and KN-AdS BH \cite{jh} one can not find a simple area 
product relation, instead one could find a complicated function of event horizon(${\cal H}^{+}$) 
area and Cauchy horizon area (${\cal H}^{-}$) that might be universal. Very recently, we derived 
for a regular Ay\'{o}n-Beato and Garc\'{i}a BH (ABG) \cite{ppgrg} the function of ${\cal H}^{+}$ 
area and ${\cal H}^{-}$ area is 
\begin{eqnarray}
 f({\cal A}_{+},{\cal A}_{-}) &=& 384 \pi^3 q^6  ~.\label{eq}
\end{eqnarray}
where the function should read
$$
f({\cal A}_{+},{\cal A}_{-})={\cal A}_{+}{\cal A}_{-}({\cal A}_{+}+{\cal A}_{-})+24 \pi q^2 
{\cal A}_{+}{\cal A}_{-}-256 \pi ^4 q^8 (\frac{{\cal A}_{+}+{\cal A}_{-}}{{\cal A}_{+}{\cal A}_{-}})
-
$$
$$
\frac{{\cal A}_{+}{\cal A}_{+}}{{\cal A}_{+}+{\cal A}_{+} +4\pi q^2}\left[ ({\cal A}_{+}+{\cal A}_{-})^2 
+24 \pi q^2 ({\cal A}_{+}+{\cal A}_{-})-{\cal A}_{+}{\cal A}_{-}-\frac{256 \pi^4 q^8}
{{\cal A}_{+}{\cal A}_{-}} +176 \pi^2 q^4 \right]
$$
It indicates a very complicated function of horizons area that turns out to be universal. But it is not 
a simple area product of horizon radii as in RN BH, Kerr BH and KN BH. This has been very popular topic 
in recent years in the GR(General Relativity) community \cite{ah09} as well as in the String community \cite{cgp11}
(see also\cite{cr12,sd12,don,pp14,pp15}). These universal relations are particularly interesting because 
they could hold in a more general situations like when a BH space-time is perturbed by surrounding matter. 
For example, KN BH  surrounded by a ring of matter the universal relation does indeed holds \cite{ah09}.

Recently, Page et al.\cite{don} have given a heuristic argument for the universal area product relation 
of a four-dimensional adiabatically distorted charged rotating BH. They in fact showed the product of 
outer horizon area and inner horizon area could be expressed in terms of a polynomial function of its 
charge, angular momenta and inverse sqare root of cosmological constant. It has been argued by Cveti\v{c}
et al. \cite{cgp11} if the cosmological parameter is quantized, the product of ${\cal H}^{+}$ area and 
${\cal H}^{-}$ area could provide a ``looking glass for probing the microscopis of general BHs''.

However in this work, we would like to evaluate the thermodynamic product formula for a generalized 
regular(singularity free) BH described by the four parameters namely, $m$, $q$, $\alpha$ and $\beta$ \cite{abg1}.
This class of BH is a solution of Einstein equations coupled with a non-linear electrodynamics source.
We examine complete thermodynamic properties of this BH. We show that some complicated function of 
physical horizon areas that is indeed mass-independent but it is not a simple area product as in Kerr 
BH or KN BH. We also compute the specific heat to examine the thermodynamic stability of the BH. Finally we 
compute the Komar energy of this BH. 

It should be noted that Smarr's mass formula and Bose-Dadhich identity do not hold for $\alpha=3$ 
and $\beta=3$ when one has taken into account the non-linear electrodynamics \cite{breton,bose,balrat,ansoldi}.
It also should be mentioned that for some regular BHs \cite{ma}, once the entropy is taken to be the Bekenstein-Hawking entropy 
\cite{bk73,bcw73} the first law of BH thermodynamics is no longer established because there is an inconsistency between the 
conventional 1st law of BH mechanics and Bekenstein-Hawking area law. The authors \cite{ma} also showed that the corrected form 
of the first law of BH thermodynamics for these class of regular BHs.  We should expect this is also true for regular ABG BH and 
it should be valid  for arbitrary values of $\alpha$ and $\beta$.

The plan of the paper is as follows. In Sec. 2, we have described the basic properties of the 
generalized regular BH and computed various thermodynamic properties.  Finally, we conclude our discussions in Sec. 3.

%%%%%%%%%%%%%%%%%%%%%%%%%%%%%%%%%%%%%%%%%%%%%%%%%%%%%%%%%%%%%%%%%%%%%%%%%%%%%%%%%%%%%%%%
\section{Generalized regular BH solution:}
The gravitational field around the four parametric regular BH solution is described by the metric 
\begin{eqnarray}
ds^2=-{\cal B}(r)dt^{2}+\frac{dr^{2}}{{\cal B}(r)}+r^{2}(d\theta^{2}+\sin^{2}\theta d\phi^{2}) 
~.\label{ag1}
\end{eqnarray}
where the function ${\cal B}(r)$ is defined by
\begin{eqnarray}
{\cal B}(r) &=& 1-\frac{2mr^{\alpha-1}}{(r^2+q^2)^{\frac{\alpha}{2}}}+\frac{q^2r^{\beta-2}}
{(r^2+q^2)^{\frac{\beta}{2}}} ~.\label{ag2}
\end{eqnarray}
and the electric field is given by 
\begin{eqnarray}
{\cal E}(r) &=& q\left[\frac{m\alpha\{5r^2-(\alpha-3)q^2\}r^{\alpha-1}}{(r^2+q^2)^{\frac{\alpha}{2}+2}}
+\frac{\{4r^4-(7\beta-8)q^2r^2+(\beta-1)(\beta-4)^4\}r^{\frac{\beta}{2}}}{4(r^2+q^2)^{\frac{\beta}{2}+2}} \right]  ~.\label{abg3}
\end{eqnarray}
where the parameters are described previously. This is a class of regular(curvature free) BH 
solution in GR. The source is a non-linear electrodynamics. In the weak field limits the 
non-linear electrodynamics becomes Maxwell field. 

In the asymptotic limit, the above solution behaves as 
\begin{eqnarray}
-g_{tt} &=& 1-\frac{2m}{r}+\frac{q^2}{r^2}+\alpha\frac{m q^2}{r^3}-\beta\frac{q^4}{2r^4}
+{\cal O}\left(\frac{1}{r^5}\right), \\
{\cal E}(r) &=& \frac{q}{r^2} +\alpha\frac{5m}{2r^3}-\beta\frac{9q^3}{4r^4}+{\cal O}
\left(\frac{1}{r^5}\right) ~.\label{ag3}
\end{eqnarray}
It may be noted that up to ${\cal O}\left(\frac{1}{r^3}\right)$, we recover the charged BH 
behavior and the parameters $m$ and $q$ are related to the mass and charges respectively. 
From the electric field behavior we can say that $\alpha$ and $\beta$ are associated to the 
electric dipole moment and electric quadrupole moments respectively.
It also should be noted that in the limit $\alpha=\beta=0$, we obtain the RN BH and in the limit 
$\alpha=3$ and $\beta=4$, we recover the ABG BH. This BH solution can be treated as a 
generalization of ABG BH \cite{abg}. The above metric as well as  the scalar invariants 
$R, R_{ab}R^{ab}, R_{abcd}R^{abcd}$ and the electric field are regular everywhere in the 
space-time. Hence, in this sense it is called a regular BH in Einstein-Maxwell gravity. The first 
regular model is discovered by Bardeen \cite{bard} in 1968.

The BH horizons can be obtain by setting ${\cal B}(r)=0$ i.e. 
\begin{eqnarray} 
(r^2+q^2)^{\frac{\alpha+\beta}{2}}-2mr^{\alpha-1}(r^2+q^2)^{\frac{\beta}{2}}
+q^2r^{\beta-2}(r^2+q^2)^{\frac{\alpha}{2}} &=& 0 . ~\label{ag4} 
\end{eqnarray}
In this work we restrict our case $\alpha\geq 3$ and $\beta \geq 4$. 

{\centerline{\bf Case-I}}
We have set $\alpha=4$ and $\beta=5$. In this case the horizon 
equation is found to be 
$$
r^{10}-2mr^9+(4m^2+5q^2)r^{8}-6mq^2r^7+(4m^2q^2+9q^4)r^{6}-6mq^4r^{5}+
$$
\begin{eqnarray}
10q^6r^4-2mq^6r^3+5q^8r^2+q^{10} &=& 0 ~.\label{ag5}
\end{eqnarray}
To finding the roots we apply the Vieta's theorem, we find
\begin{eqnarray}
\sum_{i=1}^{10} r_{i} &=& 2m~.\label{eq1}\\
\sum_{1\leq i<j\leq 10} r_{i}r_{j} &=& 4m^2+5q^2 ~.\label{eq2}\\
\sum_{1\leq i<j<k\leq 10} r_{i}r_{j} r_{k} &=& 6mq^2 ~.\label{eq3}\\
\sum_{1\leq i<j<k<l\leq 10} r_{i}r_{j} r_{k}r_{l} &=& 4m^2q^2+9q^4  ~.\label{eq4}\\
\sum_{1\leq i<j<k<l<p\leq 10} r_{i}r_{j} r_{k}r_{l}r_{p} &=& 6mq^4  ~.\label{eq5}\\
\sum_{1\leq i<j<k<l<p<s\leq 10} r_{i}r_{j} r_{k}r_{l}r_{p}r_{s} &=& 10q^6  ~.\label{eq6}\\
\sum_{1\leq i<j<k<l<p<s<t\leq 10} r_{i}r_{j} r_{k}r_{l}r_{p}r_{s}r_{t} &=& 2mq^6  ~.\label{eq7}\\
\sum_{1\leq i<j<k<l<p<s<t<u\leq 10} r_{i}r_{j} r_{k}r_{l}r_{p}r_{s}r_{t}r_{u} &=& 5q^8  ~.\label{eq8}\\
\prod_{i=1}^{10}r_{i} &=& q^{10}
\end{eqnarray}
Eliminating mass parameter we find the following set of equations:
\begin{eqnarray}
\sum_{1\leq i<j\leq 10} r_{i}r_{j}-\left(\sum_{i=1}^{10} r_{i}\right)^2 &=& 5q^2 ~.\label{eq9}\\
\sum_{1\leq i<j<k\leq 10} r_{i}r_{j} r_{k} &=& 3q^2 \sum_{i=1}^{10} r_{i} ~.\label{eq10}\\
\sum_{1\leq i<j<k<l\leq 10} r_{i}r_{j} r_{k}r_{l}-q^2\left(\sum_{i=1}^{10} r_{i}\right)^2 
&=& 9q^4  ~.\label{eq11}\\
\sum_{1\leq i<j<k<l<p\leq 10} r_{i}r_{j} r_{k}r_{l}r_{p} &=& 3q^4\sum_{i=1}^{10} r_{i} ~.\label{eq12}\\
\sum_{1\leq i<j<k<l<p<s<t\leq 10} r_{i}r_{j} r_{k}r_{l}r_{p}r_{s}r_{t}
&=& q^6 \sum_{i=1}^{10} r_{i} ~.\label{eq13}\\
\end{eqnarray}
Eliminating further finally we find the following mass independent relations: 
\begin{eqnarray}
\sum_{1\leq i<j<k<l\leq 10} r_{i}r_{j} r_{k}r_{l}-q^2\sum_{1\leq i<j\leq 10} r_{i}r_{j}
&=& 4q^4  ~.\label{eq14}\\
3\sum_{1\leq i<j<k<l<p<s<t\leq 10} r_{i}r_{j} r_{k}r_{l}r_{p}r_{s}r_{t} &=& 
q^2 \sum_{1\leq i<j<k<l<p\leq 10} r_{i}r_{j} r_{k}r_{l}r_{p}~.\label{eq15}\\
\sum_{1\leq i<j<k<l<p<s\leq 10} r_{i}r_{j} r_{k}r_{l}r_{p}r_{s} &=& 10q^6 ~.\label{eq16}\\
\sum_{1\leq i<j<k<l<p<s<t<u\leq 10} r_{i}r_{j} r_{k}r_{l}r_{p}r_{s}r_{t}r_{u} &=& 5q^8  
~.\label{eq17}\\
\prod_{i=1}^{10}r_{i} &=& q^{10}
\end{eqnarray}
In terms of area ${\cal A}_{i}=4 \pi r_{i}^2$, the mass-independent relations are
\begin{eqnarray}
\sum_{1\leq i<j<k<l\leq 10} \sqrt{{\cal A}_{i}{\cal A}_{j} {\cal A}_{k}{\cal A}_{l}}
-4\pi q^2\sum_{1\leq i<j\leq 10} \sqrt{{\cal A}_{i}{\cal A}_{j}}
&=& (8\pi q^2)^2  ~.\label{eq18}\\
3\sum_{1\leq i<j<k<l<p<s<t\leq 10} \sqrt{{\cal A}_{i}{\cal A}_{j} {\cal A}_{k}{\cal A}_{l}{\cal A}_{p}
{\cal A}_{s}{\cal A}_{t}} &=& 
4\pi q^2 \sum_{1\leq i<j<k<l<p\leq 10} \sqrt{{\cal A}_{i}{\cal A}_{j} {\cal A}_{k}{\cal A}_{l}
{\cal A}_{p}}~.\label{eq19}\nonumber\\
\sum_{1\leq i<j<k<l<p<s\leq 10} \sqrt{{\cal A}_{i}{\cal A}_{j} {\cal A}_{k}{\cal A}_{l}{\cal A}_{p}
{\cal A}_{s}} &=& 640\pi^3q^6 ~.\label{eq20}\\
\sum_{1\leq i<j<k<l<p<s<t<u\leq 10} \sqrt{{\cal A}_{i}{\cal A}_{j} {\cal A}_{k}{\cal A}_{l}{\cal A}_{p}
{\cal A}_{s}{\cal A}_{t}{\cal A}_{u}} &=& 1280 \pi^4 q^8  
~.\label{eq21}\\
\prod_{i=1}^{10}\sqrt{{\cal A}_{i}} &=& (4\pi q^{2})^5 ~.\label{eq22}
\end{eqnarray}
From this relation one can obtain the mass-independent entropy relation by substituting 
${\cal A}_{i}=4{\cal S}_{i}$. Thus the mass-independent complicated function of horizon
areas that we have found could turn out to be an \emph{universal} quantity.

{\centerline{\bf Case-II}}
Now we have set $\alpha=5$ and $\beta=6$. In this case the horizon 
equation is given by 
\begin{eqnarray}
r^{12}-(4m^2-8q^2)r^{10}-(4m^2q^2-22q^4)r^{8}+26q^6r^{6}+17q^8r^4+6q^{10}r^2+q^{12} 
&=& 0 ~.\label{ag6}\nonumber
\end{eqnarray}
Let us put $r^2=x$ then the equation reduces to sixth order polynomial equation:
\begin{eqnarray}
x^{6}-(4m^2-8q^2)x^{5}-(4m^2q^2-22q^4)x^{4}+26q^6x^{3}+17q^8x^2+6q^{10}x+q^{12} 
&=& 0 ~.\label{ag7}\nonumber
\end{eqnarray}
Again we apply the Vieta's theorem, we get
\begin{eqnarray}
\sum_{i=1}^{6} x_{i} &=& 4m^2-8q^2 ~.\label{ag8}\\
\sum_{1\leq i<j\leq 6} x_{i}x_{j} &=& 22q^4-4m^2q^2  ~.\label{ag9}\\
\sum_{1\leq i<j<k\leq 6} x_{i}x_{j} x_{k} &=& -26q^6 ~.\label{ag10}\\
\sum_{1\leq i<j<k<l\leq 6} x_{i}x_{j} x_{k}x_{l} &=& 17q^8  ~.\label{ag11}\\
\sum_{1\leq i<j<k<l<p\leq 6} x_{i}x_{j}x_{k}x_{l}x_{p} &=& -6q^{10}  ~.\label{ag12}\\
\prod_{i=1}^{6}x_{i} &=& q^{12} ~.\label{ag13}
\end{eqnarray} 
Eliminating mass parameter, we obtain the mass-independent relation:
\begin{eqnarray}
\sum_{1\leq i<j\leq 6} x_{i}x_{j}+q^2 \sum_{i=1}^{6} x_{i}  &=& 14q^{4}  ~.\label{ag14}\\
\sum_{1\leq i<j<k\leq 6} x_{i}x_{j} x_{k} &=& -26q^6 ~.\label{ag15}\\
\sum_{1\leq i<j<k<l\leq 6} x_{i}x_{j} x_{k}x_{l} &=& 17q^8  ~.\label{ag16}\\
\sum_{1\leq i<j<k<l<p\leq 6} x_{i}x_{j}x_{k}x_{l}x_{p} &=& -6q^{10}  ~.\label{ag17}\\
\prod_{i=1}^{6}x_{i} &=& q^{12} ~.\label{ag18}
\end{eqnarray} 
In terms of area ${\cal A}_{i}=4\pi r_{i}^2=4 \pi x_{i}$ the above mass-independent relation 
could be written as 
\begin{eqnarray}
\sum_{1\leq i<j\leq 6} {\cal A}_{i}{\cal A}_{j}+4\pi q^2 \sum_{i=1}^{6} {\cal A}_{i}  
&=& 14(4\pi q^{2})^2  ~.\label{ag19}\\
\sum_{1\leq i<j<k\leq 6} {\cal A}_{i}{\cal A}_{j} {\cal A}_{k} &=& -26(4\pi q^2)^3 ~.\label{ag20}\\
\sum_{1\leq i<j<k<l\leq 6} {\cal A}_{i}{\cal A}_{j} {\cal A}_{k}{\cal A}_{l} &=& 
17(4\pi q^2)^4  ~.\label{ag21}\\
\sum_{1\leq i<j<k<l<p\leq 6} {\cal A}_{i}{\cal A}_{j}{\cal A}_{k}{\cal A}_{l}{\cal A}_{p} 
&=&-6(4\pi q^{2})^5  ~.\label{ag22}\\
\prod_{i=1}^{6}{\cal A}_{i} &=& (4\pi q^{2})^6 ~.\label{ag23}
\end{eqnarray} 
Again we have found mass-independent complicated function of horizon areas that could 
turn out to be universal in nature.
 
{\centerline{\bf Case-III}}
Now we have set $\alpha=4$ and $\beta=6$. In this case the horizon 
equation is
$$
r^{10}-2mr^9+6q^2r^{8}-6mq^2r^7+12q^4r^{6}-6mq^4r^{5}+
$$
\begin{eqnarray}
11q^6r^4-2mq^6r^3+5q^8r^2+q^{10} &=& 0 ~.\label{a1}
\end{eqnarray}

Again we apply the Vieta's theorem, we get
\begin{eqnarray}
\sum_{i=1}^{10} r_{i} &=& 2m~.\label{eqq1}\\
\sum_{1\leq i<j\leq 10} r_{i}r_{j} &=& 6q^2 ~.\label{eqq2}\\
\sum_{1\leq i<j<k\leq 10} r_{i}r_{j} r_{k} &=& 6mq^2 ~.\label{eqq3}\\
\sum_{1\leq i<j<k<l\leq 10} r_{i}r_{j} r_{k}r_{l} &=& 12q^4  ~.\label{eqq4}\\
\sum_{1\leq i<j<k<l<p\leq 10} r_{i}r_{j} r_{k}r_{l}r_{p} &=& 6mq^4  ~.\label{eqq5}\\
\sum_{1\leq i<j<k<l<p<s\leq 10} r_{i}r_{j} r_{k}r_{l}r_{p}r_{s} &=& 11q^6  ~.\label{eqq6}\\
\sum_{1\leq i<j<k<l<p<s<t\leq 10} r_{i}r_{j} r_{k}r_{l}r_{p}r_{s}r_{t} &=& 2mq^6  ~.\label{eqq7}\\
\sum_{1\leq i<j<k<l<p<s<t<u\leq 10} r_{i}r_{j} r_{k}r_{l}r_{p}r_{s}r_{t}r_{u} &=& 5q^8  ~.\label{eqq8}\\
\prod_{i=1}^{10}r_{i} &=& q^{10} ~.\label{eqq9}
\end{eqnarray}
Eliminating mass parameter we have found the following set of mass-independent relation:
\begin{eqnarray}
3\sum_{1\leq i<j<k<l<p<s<t\leq 10} r_{i}r_{j} r_{k}r_{l}r_{p}r_{s}r_{t} &=& 
q^2 \sum_{1\leq i<j<k<l<p\leq 10} r_{i}r_{j} r_{k}r_{l}r_{p}~.\label{eqq10}\\
\sum_{1\leq i<j\leq 10} r_{i}r_{j} &=& 6q^2 ~.\label{eqq11}\\
\sum_{1\leq i<j<k<l\leq 10} r_{i}r_{j} r_{k}r_{l} &=& 12q^4  ~.\label{eqq12}\\
\sum_{1\leq i<j<k<l<p<s\leq 10} r_{i}r_{j} r_{k}r_{l}r_{p}r_{s} &=& 11q^6  ~.\label{eqq13}\\
\sum_{1\leq i<j<k<l<p<s<t<u\leq 10} r_{i}r_{j} r_{k}r_{l}r_{p}r_{s}r_{t}r_{u} &=& 5q^8  ~.\label{eqq14}\\
\prod_{i=1}^{10}r_{i} &=& q^{10} ~.\label{eqq15}
\end{eqnarray}
If we are working in terms of area then the above mass-independent relation could be written as 
\begin{eqnarray}
3\sum_{1\leq i<j<k<l<p<s<t\leq 10} \sqrt{{\cal A}_{i}{\cal A}_{j} {\cal A}_{k}{\cal A}_{l}{\cal A}_{p}
{\cal A}_{s}{\cal A}_{t}} &=& 
4\pi q^2 \sum_{1\leq i<j<k<l<p\leq 10} \sqrt{{\cal A}_{i}{\cal A}_{j} {\cal A}_{k}{\cal A}_{l}
{\cal A}_{p}}~.\label{eqq16}\nonumber\\
\sum_{1\leq i<j\leq 10} \sqrt{{\cal A}_{i}{\cal A}_{j}} &=& 24\pi q^2 ~.\label{eqq17}\\
\sum_{1\leq i<j<k<l\leq 10} \sqrt{{\cal A}_{i}{\cal A}_{j} {\cal A}_{k}{\cal A}_{l}} &=& 
3(8\pi q^2)^2  ~.\label{eqq18}\\
\sum_{1\leq i<j<k<l<p<s\leq 10} \sqrt{{\cal A}_{i}{\cal A}_{j} {\cal A}_{k}{\cal A}_{l}{\cal A}_{p}{\cal A}_{s}} &=& 11(4\pi q^2)^3  ~.\label{eqq19}\\
\sum_{1\leq i<j<k<l<p<s<t<u\leq 10} \sqrt{{\cal A}_{i}{\cal A}_{j} {\cal A}_{k}{\cal A}_{l}{\cal A}_{p}{\cal A}_{s}{\cal A}_{t}{\cal A}_{u}} &=& 5(4\pi q^2)^4  ~.\label{eqq20}\\
\prod_{i=1}^{10}\sqrt{{\cal A}_{i}} &=& (4\pi q^{2})^5 ~.\label{eqq21}
\end{eqnarray}
Once again these are mass-independent relations that could turn out to be universal. So on, we can compute 
different thermodynamic product relations for different values of $\alpha$ and $\beta$ which 
are mass-independent.

The Hawking \cite{bcw73} temperature on ${\cal H}^{i}$ reads off
\begin{eqnarray}
T_{i} &=& \frac{\kappa_{i}}{2\pi}=\frac{1}{4\pi}
\left[\alpha\frac{r_{i}}{(r_{i}^2+q^2)}-\frac{(\alpha-1)}{r_{i}}+
\frac{q^2(\alpha-\beta)r_{i}^{\beta-1}}{(r_{i}^2+q^2)^{\frac{\beta}{2}+1}}+
 \frac{q^2(\beta-\alpha-1)r_{i}^{\beta-2}}{(r_{i}^2+q^2)^{\frac{\beta}{2}}}
\right] .~\label{agg}
\end{eqnarray}
It should be noted that the Hawking temperature product is depend on the mass parameter 
and thus it is not an universal quantity.

Another important parameter in BH thermodynamics that determine the thermodynamic stability 
of BH is defined by 
\begin{eqnarray}
C_{i} &=& \frac{\partial m}{\partial T_{i}} .~\label{ag25}
\end{eqnarray}
For generalized regular BH, it is found to be 
\begin{eqnarray}
C_{i} &=& \frac{\left(\frac{\partial m}{\partial r_{i}}\right)}
{\left(\frac{\partial T_{i}}{\partial r_{i}}\right)} ~\label{ag26}
\end{eqnarray}
where 
$$
\frac{\partial m}{\partial r_{i}} = \frac{\left[\alpha r_{i}^{\alpha}(r_{i}^2+q^2)^{\frac{\alpha}{2}-1}
-(\alpha-1)r_{i}^{\alpha-2}(r_{i}^2+q^2)^{\frac{\alpha}{2}}\right]}{2r_{i}^{2(\alpha-1)}}+
$$
\begin{eqnarray}
\frac{q^2\left[(\beta-\alpha-1)r_{i}^{\beta-\alpha-2}(r_{i}^2+q^2)^{\frac{\beta-\alpha}{2}}
-(\beta-\alpha)r_{i}^{\beta-\alpha}(r_{i}^2+q^2)^{\frac{\beta-\alpha-2}{2}}\right]}
{(r_{i}^2+q^2)^{\beta-\alpha}}
~\label{ag27}
\end{eqnarray}
and
$$
\frac{\partial T_{i}}{\partial r_{i}} =\frac{1}{4\pi}\left[\frac{\alpha-1}{r_{i}^2}
-\frac{\alpha (r_{i}^2-q^2)}{(r_{i}^2+q^2)^2}+q^2(\alpha-\beta)
\frac{\{(\beta-1)r_{i}^{\beta-2}(r_{i}^2+q^2)^{\frac{\beta}{2}+1}-(\beta+2)r_{i}^{\beta}
(r_{i}^2+q^2)^{\frac{\beta}{2}}\}}{(r_{i}^2+q^2)^{\beta+2}}\right]+
$$
\begin{eqnarray}
\frac{1}{4\pi}\left[q^2(\beta-\alpha-1)\frac{\{(\beta-2)r_{i}^{\beta-3}(r_{i}^2+q^2)^{\frac{\beta}{2}}
-\beta r_{i}^{\beta-1}(r_{i}^2+q^2)^{\frac{\beta}{2}-1}\}}{(r_{i}^2+q^2)^{\beta}} \right]
~\label{ag28}
\end{eqnarray}
The BH undergoes a second order phase transition when $\frac{\partial T_{i}}{\partial r_{i}}=0$. In this case 
the specific heat diverges.

Finally, the Komar \cite{ak59} energy computed at ${\cal H}^{i}$ is given by
\begin{eqnarray}
E_{i} &=& \frac{1}{2} \left[\alpha\frac{r_{i}^3}{(r_{i}^2+q^2)}-(\alpha-1)r_{i}+
\frac{q^2(\alpha-\beta)r_{i}^{\beta+1}}{(r_{i}^2+q^2)^{\frac{\beta}{2}+1}}+
 \frac{q^2(\beta-\alpha-1)r_{i}^{\beta}}{(r_{i}^2+q^2)^{\frac{\beta}{2}}} \right] 
.\label{ag29}
\end{eqnarray}
In the limit $\alpha=3$ and $\beta=4$, one obtains the result of ABG BH.

\section{Discussion:}
In this work, we examined thermodynamic product relations for generalized 
regular(curvature free) BH. Generalized means the BHs represented by the four parameters i.e. 
$m$, $q$, $\alpha$ and $\beta$. We determined different thermodynamic product particularly 
area products for different values of $\alpha$ and $\beta$. We showed that there are some 
complicated function of horizon areas indeed mass-independent that could turn out to be 
universal. We also derived the specific heat to determine the thermodynamic stability of 
the BH. Finally, we computed the Komar energy for this generalized BH.

%\section*{References}

\end{document}